\documentclass[english,aps,prl,showpacs,twocolumn]{revtex4}
\usepackage[T1]{fontenc}
\usepackage[latin1]{inputenc}
\usepackage{graphicx}

\makeatletter

\providecommand{\tabularnewline}{\\}

\usepackage{babel}
\makeatother
\begin{document}

\preprint{This line only printed with preprint option}

\title{Phase synchronization in cerebral hemodynamics}

\author{Miroslaw Latka}

\email{Miroslaw.Latka@pwr.wroc.pl}

\homepage{http://www.if.pwr.wroc.pl/~mirek/}

\affiliation{Institute of Physics, Wroclaw University of Technology, Wybrzeze
Wyspianskiego 27, 53-227 Wroclaw, Poland}

\author{Malgorzata Turalska}

\affiliation{Institute of Physics, Wroclaw University of Technology, Wybrzeze
Wyspianskiego 27, 53-227 Wroclaw, Poland}

\author{Waldemar Kolodziej}

\email{wkolodziej@wcm.opole.pl}

\affiliation{Department of Neurosurgury, Opole Regional Medical Center, Al. Witosa
26, 45-401 Opole, Poland}

\author{Dariusz Latka}

\affiliation{Department of Neurosurgury, Opole Regional Medical Center, Al. Witosa
26, 45-401 Opole, Poland}

\email{dlatka@wcm.opole.pl}

\author{Brahm Goldstein}

\email{goldsteb@ohsu.edu}

\homepage{www.ohsuhealth.com/dch/complex}

\affiliation{Doernbecher Children's Hospital, Oregon Health \& Science University,
707 SW Gaines St. Mail Code CDRCP Portland, OR 97239 }

\author{Bruce J. West}

\email{Bruce.J.West@us.army.mil}

\affiliation{Mathematics Division, Army Research Office, P.O. Box 12211, Research
Triangle, NC 27709-2211, USA}

\date{June 8, 2004}

\begin{abstract}
A healthy human brain is perfused with blood flowing laminarly through
cerebral vessels, providing brain tissue with substrates such as oxygen
and glucose. Under normal conditions, cerebral blood flow is controlled
by autoregulation as well as metabolic, chemical and neurogenic regulation.
Physiological complexity of these mechanisms invariably leads to a
question as to what are the relations between the statistical properties
of arterial and intracranial pressure fluctuations. To shed new light
on cerebral hemodynamics, we employ a complex continuous wavelet transform
to determine the instantaneous phase difference between the arterial
blood pressure (ABP) and intracranial pressure (ICP) in patients with
traumatic brain injuries or spontaneous cerebral hemorrhage. For patients
with mild to moderate injury, the phase difference \emph{slowly} evolves
in time. However, severe neurological injury with elevated ICP are
herein associated with \emph{synchronization} of arterial and intracranial
pressure. We use Shannon entropy to quantify the stability of ABP-ICP
phase difference and discuss the clinical applicability of such measure
to assessment of cerebrovascular reactivity and autoregulation integrity.
\end{abstract}


\maketitle
Synchronization between different physiological systems or subsystems
has long been recognized as a ubiquitous dynamical effect \cite{glass88}.
Examples of such synchronization are as diverse as circadian rhythm
\cite{glass88,aschoff82}, correlation of respiration with mechanical
ventilation \cite{petrillo84} or locomotory rhythm \cite{bramble83},
coordinated motion \cite{glass88}, animal gait \cite{collins93},
or synchronization of oscillations of human insulin secretion with
glucose infusion \cite{sturis95}. These phenomena are essentially
confined to nearly periodic rhythms. However, the development of novel
concepts \cite{tass98,rosenblum98} has paved the way for the application
of synchrony analysis to inherently nonstationary and noisy signals;
time series that are characteristic of cardiology and encephalography
\cite{schafer98,quyen01} (see also references therein).

A healthy human brain is perfused with blood flowing laminarly through
cerebral vessels, providing brain tissue with substrates such as oxygen
and glucose. Cerebral blood flow (CBF) is relatively stable, with
typical values between 45 and 65 ml/100g of brain tissue per second,
despite variations in systemic pressure as large as 100 Torr. This
phenomenon, known as cerebral autoregulation (CA) \cite{paulson90},
is mainly associated with changes in cerebrovascular resistance of
small precapillary brain arteries. CBF is also affected by metabolic,
chemical and neurogenic regulation. Strong susceptibility of brain
tissue to even short periods of ischemia underlies the physiological
significance of these intricate control mechanisms. In the phenomenological
description of cerebral hemodynamics, fluctuations of arterial blood
pressure (ABP), due to pressure reactivity of cerebral vessels, lead
to fluctuations of intracranial pressure (ICP). The goal of this paper
is to quantitatively analyze the interplay of ABP and ICP from the
viewpoint of synchronization. In particular, we investigate time evolution
of instantaneous phase difference between ABP and ICP time series
in patients with traumatic brain injuries or spontaneous cerebral
hemorrhage. We examine the extent to which the relative phase dynamics
reflects pathological conditions. We adopt the mathematical framework
of synchronization since it is particularly well-suited to the analysis
of non-stationary bivariate time series. Interestingly enough, we
have not been able to find previous application of this approach to
cerebral hemodynamics, see, for example \cite{Aaslid89,Diehl95,Tiecks95,Panerai99,Zhang98,czosnyka02,czosnyka03,balestreri04}.

Let us consider two signals $s_{1}(t)$, $s_{2}(t)$ and their corresponding
instantaneous phases $\phi_{1}(t)$ and $\phi_{2}(t)$. Phase synchronization
takes place when $n\phi_{1}(t)-m\phi_{2}(t)=const$ where $n$, $m$
are integers indicating the ratios of possible frequency locking.
Herein we consider only the simplest case $n=m=1.$ Furthermore, as
with most biological signals contaminated by noise, we are forced
to search for approximate phase synchrony, i.e. $\phi_{1}(t)-\phi_{2}(t)\approx const$.
Thus, the studies of synchronization involve not only the determination
of instantaneous phases of signals but also the introduction of some
statistical measure of phase locking.

The wavelet transform is an integral transform for which basis functions,
known as wavelets, are well localized both in time and frequency \cite{mallat98}.
Moreover, the wavelet basis can be constructed from a single function
$\psi(t)$ by means of translation and dilation $\psi(a;t)=\psi(t-t_{0}/a)$.
The function $\psi(t)$ is commonly referred to as the mother function
or analyzing wavelet. The wavelet transform of function $s(t)$ is
defined as 

\begin{equation}
W[s](a,t_{0})=\frac{1}{\sqrt{a}}\int_{-\infty}^{\infty}s(t)\psi^{*}(a;t_{0})dt,\label{wtransform}\end{equation}
where $\psi^{*}(t)$ denotes the complex conjugate of $\psi(t)$.
In this work we employ the Morlet wavelet:

\begin{equation}
\psi(t)=\sqrt{\pi f_{b}}e^{2\pi if_{c}t}e^{-t^{2}/f_{b}}\label{cmor}\end{equation}
 and set the bandwith parameter $f_{b}$ as well as the center frequency
$f_{c}$ to 1. For a given sampling period $\delta t$, it is possible
to associate a pseudofrequency with scale $a$:

\begin{equation}
f_{a}=\frac{\delta tf_{c}}{a}.\label{pseudo}\end{equation}
Obviously, the dual localization of wavelets makes the above frequency
assignment approximate.

The instantaneous phase $\phi(t_{0})$ of a signal $s(t)$ can be
readily extracted by calculating a wavelet transform with a complex
mother function \cite{lachaux99,quyen01}:

\begin{equation}
exp[i\phi(a,t_{0})]=W[s](a,t_{0})/\left|W[s](a,t_{0})\right|,\label{iphase}\end{equation}
where we explicitly indicated the dependence of phase on the scale
$a$ to emphasize that we are investigating frequency-specific synchronization,
\emph{i.e.} transient phase-locking.

Following Tass \emph{et al.} \cite{tass98} we characterize the strength
of phase synchronization with the help of the index $\gamma$: 

\begin{equation}
\gamma=(H_{max}-H)/H_{max},\label{gamma}\end{equation}
derived from the Shannon entropy $H$. In the well known formula $H=\sum_{k=1}^{N}p_{k}lnp_{k}$,
$N$ is the number of bins and $p_{k}$ is the relative frequency
of finding the phase difference within the $k$-th bin. Due to normalization
in (\ref{gamma}), the synchronization index lies in the unit interval
$0\leq\gamma\leq1$. A vanishing index $\gamma=0$ corresponds to
uniform distribution of phase differences (no synchronization) while
$\gamma=1$ corresponds to perfect synchronization (phase locking
of the two processes).

We have applied the synchronization theory formalism (\emph{cf.} equations
(\ref{iphase}) and (\ref{gamma})) to analyze the instantaneous phase
difference between ICP and ABP time series. Both pressures were \emph{averaged}
over a cardiac cycle. The hemodynamic time series were invasively
acquired during long-term monitoring of patients with traumatic brain
injuries or spontaneous cerebral hemorrhage. The study comprised 10
juvenile patients who were admitted to the Pediatric Intensive Care
Unit of Doernbecher Children's Hospital and 10 adult subjects who
underwent the surgery at the Department of Neurosurgery of Opole Regional
Medical Center.

In Table \ref{Data} we collected the values of mean arterial and
intracranial pressures for a juvenile patient with traumatic brain
injury. From two long-term monitoring sessions, we have chosen 30
min segments from the time series and labeled them $A$, $B$ and
$C$.

\begin{figure}
\includegraphics{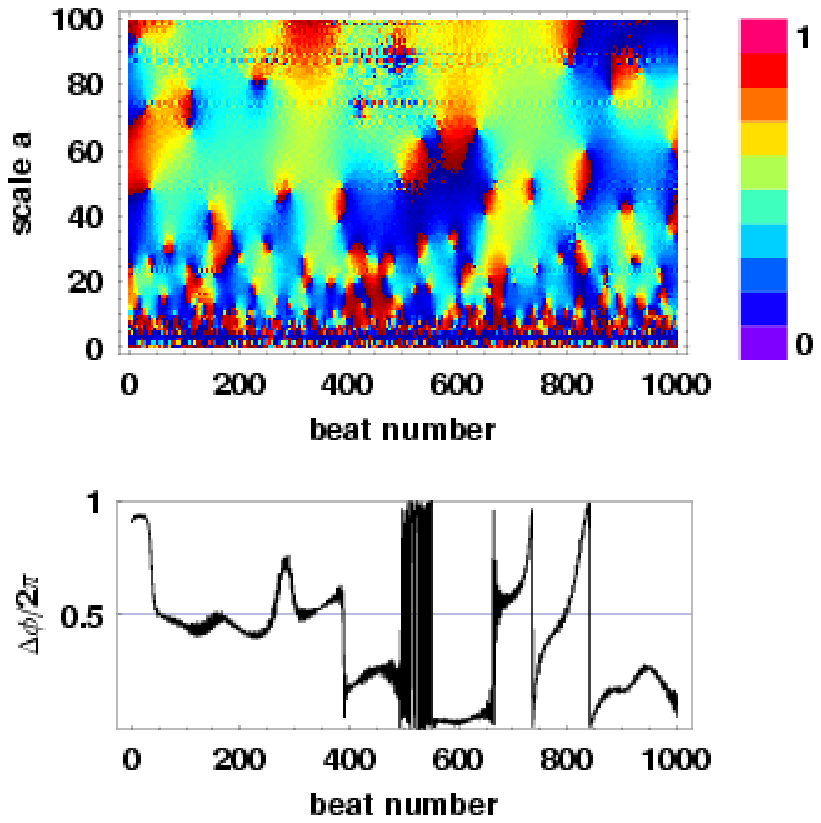}\includegraphics{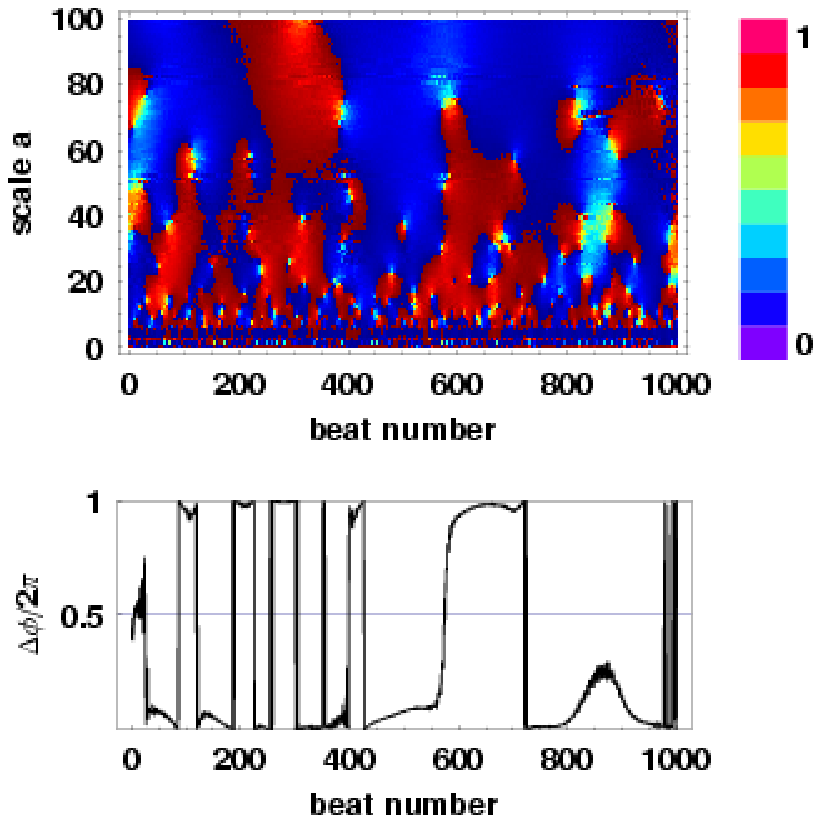}

\caption{\label{Phase15} Normalized instantaneous phase difference between
the arterial and intracranial pressure calculated with the Morlet
wavelet for 100 integer values of scale $a$. Left column corresponds
to segment \emph{A} in the first monitoring session, right column
to segment \emph{B} in the second (\emph{cf.} Table \ref{Data}).
Bottom panels show the time evolution of phases for $a=50$.}
\end{figure}

In the left column of Fig. \ref{Phase15} we present the synchrony
analysis for data segment \emph{A} from the first session. During
this period the intracranial pressure remained at an elevated but
physiologically acceptable level. It is interesting that for larger
scales the phase difference between ICP and ABP evolves very slowly.
Moreover, the distribution of colors indicates that the normalized
phase difference fluctuates most of the time around 0.5. This behavior
is in sharp contrast with the phase dynamics for segment \emph{B}
from the second session (\emph{cf.} right column of Fig. \ref{Phase15}).
During this second period the increase in ICP was accompanied by the
drop in ABP, which is strong evidence of the intermittent failure
of regulatory mechanisms. This failure resulted in insufficient perfusion
pressure. Essentially bichromatic structure of right-top panel in
Fig. \ref{Phase15} is a clear indication of strong synchronization
between arterial and intracranial pressure. This observation is corroborated
by the plot of phase for $a=50$ which elucidates that the phase variability
merely amounts to rapid transitions between 0 and 1, equivalent values
from the point of view of synchronization.

\begin{figure}
\includegraphics{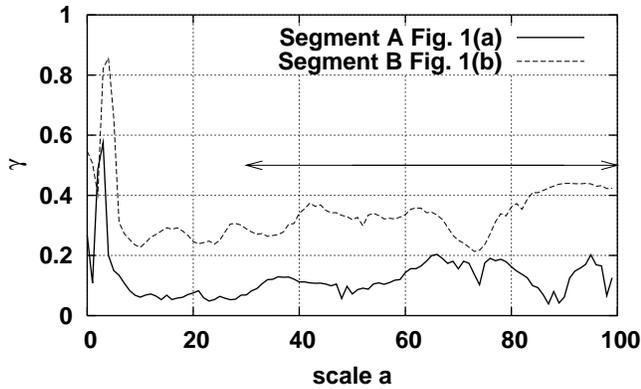}

\caption{\label{gammaExp} Synchronization parameter $\gamma$ as a function
of the scale $a$ of the complex Morlet wavelet transform. Solid line
corresponds to segment \emph{A} in the first monitoring session, dashed
line to segment \emph{B} in the second. In Table \ref{Data} we present
the value of $\gamma$ averaged over scales 30 to 100 ( as indicated
in this graph by the horizontal arrow).}
\end{figure}

In Fig. \ref{gammaExp} the synchronization parameter $\gamma$ for
the hemodynamic data analyzed in Fig. \ref{Phase15} is drawn as a
function of scale $a$ of the complex Morlet wavelet transform. For
both curves in this plot, the prominent peak is followed by a plateau.
Please note that for the chosen values of the Morlet wavelet parameters
and $\delta t=1s$ (mean cardiac interbeat interval), $1/a$ gives
the approximation of the pseudofrequency of the wavelet basis functions
$\psi(a;t_{0})$, \emph{cf.} (\ref{pseudo}). The appearance of such
high peaks in the low $a$/high pseudofrequency region is not surprising
since these maxima are merely the manifestation of the inability of
cerebral vessels to respond to rapid changes in arterial blood pressure.
In patients with severely restricted cerebral blood flow, the characteristic
peaks are often missing. In Table \ref{Data} we present the values
of the synchronization strength averaged over scales 30 to 100: $\gamma_{30:100}$.
The maximum value of 0.34 corresponds, as expected, to the high pressure/low
perfusion episode. To assess statistical significance of the observed
increase we calculated $\gamma_{30:100}$ for 49 surrogate hemodynamic
time series \cite{hegger99}. For the surrogate data we found the
mean of $\gamma_{30:100}$ equal to 0.07 and since the maximum value
was 0.13, we can reject the null hypothesis that the increase was
accidental, at the 97.5\% level of confidence.

\begin{figure}
\includegraphics{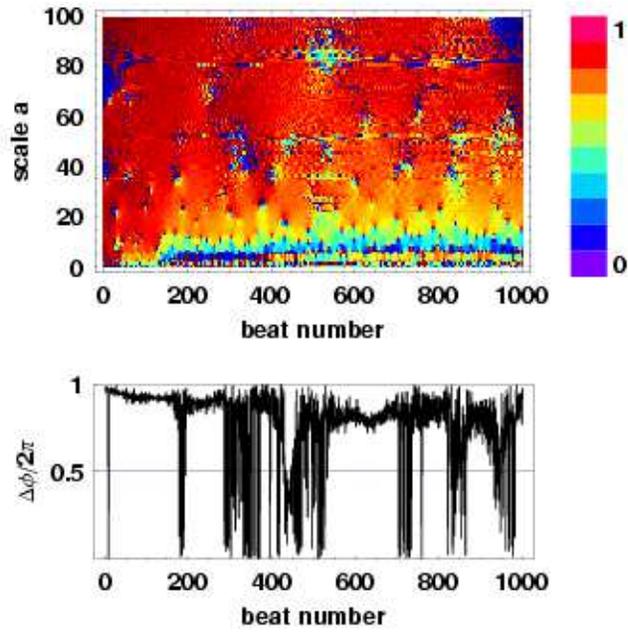}

\caption{\label{Phasebb} Normalized instantaneous phase difference between
the arterial and intracranial pressure for a patient with extremely
high ICP and negligible perfusion pressure. The calculations were
done in the same way as in Fig. \ref{Phase15}}
\end{figure}

Fig. \ref{Phasebb} shows the instantaneous phase calculations for
an adult patient after a second massive subarachnoid hemorrhage which
resulted in severe cerebral edema. During monitoring, the average
ABP was equal to 81 Torr and the average ICP was equal to 73 Torr.
In this case, not only was ICP extremely high but also perfusion was
negligible. The structure of the phase map (Fig. \ref{Phasebb}) and
high value of $\gamma=0.30$ reflects strong entrainment of arterial
and intracranial pressure time series. Thus, there is no doubt that
the phase synchronization observed in the two case studies is pathological.
In fact, for seven patients with a good clinical outcome the average
value of $\gamma$ was $0.09.$ On the other hand, for five patients
with severe injury (high ICP, low cerebral perfusion) and poor clinical
outcome the synchronization parameter was consistently high and varied
between 0.30 and 0.70.

The relations between the statistical properties of arterial and intracranial
pressure fluctuations are poorly understood. Progress in understanding
has undoubtedly been hindered by the physiological complexity of mechanisms
which affect intracranial hemodynamics, see \cite{czosnyka00} for
references. In the widely accepted phenomenological description of
the interaction between ICP and ABP, under normal conditions a decrease
in ABP results in the vasolidation of cerebral vessels which increases
cerebral blood volume and consequently ICP. In pathology, cerebral
vessels are non-reactive and changes in arterial blood pressure are
passively transmitted to ICP.

Steinmeier \emph{et al.} and Czosnyka \emph{et al.} \cite{steinmaier96,czosnyka97,balestreri04}
have introduced moving cross-correlation indices which quantify the
reactivity of vessels to changes in ABP. The clinical studies demonstrated
that these indices are positively correlated with the high intracranial
pressure, low admission Glasgow Coma Scale score and poor outcome
after injury. However, moving correlation or other coherence measures
based on spectral analysis cannot separate the effects of amplitude
from those of phase in the interrelations between ABP and ICP signals.
Complex wavelet analysis provides an effective tool for the investigation
of phase relations in a chosen frequency range and consequently can
shed new light on inherently nonstationary cerebral hemodynamics.

It is worth pointing out that cerebrovascular vasomotor reactivity
reflects changes in smooth muscle tone in the arterial wall in response
to changes in transmural pressure. The same mechanism underlies cerebral
autoregulation - a fundamental neuroprotective mechanism \cite{paulson90}.
In further studies we shall thoroughly test the clinical significance
of measures derived from instantaneous phase dynamics for assessment
of autoregulation integrity. In addition, we shall establish the connection
between the phase synchronization approach introduced in this paper
and those based on moving cross-correlation indices \cite{steinmaier96,czosnyka97}
or scaling properties of arterial and intracranial pressure time series
\cite{kolodziej03,latka04a}.

We gratefully acknowledge the financial support of the U.S. Army Research
Office (Grant DAAD19-03-1-0349). B. Goldstein acknowledges the partial
support from the Thrasher Research Fund.

\bibliographystyle{apsrev}
\bibliography{cerebral,SYNCHRO}

\begin{table}[b]
\begin{tabular}{|c||ccc|ccc|}
\hline 
\emph{N}&
&
1&
&
&
2&
\tabularnewline
&
\emph{A}&
\emph{B}&
\emph{C}&
\emph{A}&
\emph{B}&
\emph{C}\tabularnewline
\hline
\hline 
$ABP$&
78.9&
77.0&
74.6&
59.3&
63.3&
70.7\tabularnewline
\hline 
$ICP$&
13.9&
13.0&
13.0&
16.4&
19.8&
9.2\tabularnewline
\hline 
$\gamma_{30:100}$&
0.13&
0.08&
0.13&
0.17&
0.34&
0.13\tabularnewline
\hline
\end{tabular}

\caption{\label{Data} The values of mean arterial $ABP$ and intracranial
$ICP$ pressures along with the averaged synchronization index $\gamma_{30:100}$
for the 30 min data segments recorded during the monitoring of a juvenile
patient with traumatic brain injury. }
\end{table}

\end{document}